\documentclass[fleqn,12pt,twoside]{article}
\usepackage[headings]{espcrc1}

\readRCS
$Id: espcrc1.tex,v 1.2 2004/02/24 11:22:11 spepping Exp $
\usepackage{graphicx}
\usepackage[figuresright]{rotating}

\newcommand{\AmS}{{\protect\the\textfont2
  A\kern-.1667em\lower.5ex\hbox{M}\kern-.125emS}}

\title{Recent developments in chiral dynamics of hadrons and hadrons in a
nuclear medium}

\author{E. Oset \address[valencia] {Departamento de Fisica Teorica e IFIC, Centro Mixto
Universidad de Valencia-CSIC, Institutos de Investigacion de Paterna, Apdo
22085, 46071 Valencia, Spain},%
       ~S. Sarkar \addressmark[valencia],
       M. J. Vicente Vacas \addressmark[valencia],
       M. Kaskulov \addressmark[valencia],
       L. Roca \addressmark[valencia],
     V.K. Magas \address[barcelona]{Departament d'Estructura i Constituents de la
     Materia, Universitat de Barcelona, Spain},
      A. Ramos \addressmark[barcelona]
        and
        H. Toki \address{RCNP, Osaka University, Japan}}
       
\runtitle{1-column format camera-ready paper in \LaTeX}
\runauthor{E. Oset}

\begin{document}

\maketitle

\begin{abstract}
In this talk I present recent developments in chiral dynamics of hadrons and
hadrons in a medium addressing the following points: interaction of the octet of
pseudoscalar mesons with
the octet of baryons of the nucleon, showing recent experimental evidence on the
existence of two $\Lambda(1405)$ states, the interaction of the octet of 
pseudoscalar mesons with the decuplet of baryons of the $\Delta$, with
particular emphasis on the $\Lambda(1520)$ resonance, dynamically generated by
this interaction. Then I review the interaction of kaons in a nuclear medium and
briefly discuss the situation around the claims of deeply bound states in
nuclei.  The large renormalization of the $\Lambda(1520)$ in the nuclear medium
is shown as another example of successful application of the chiral unitary
techniques.

\end{abstract}

\section{Introduction}
Chiral theory has caught up as an important tool to deal with hadron reactions
at low and intermediate energies, accounting for the underlying basic QCD 
dynamics. By introducing chiral Lagrangians where the explicit fields are
mesons and baryons, a perturbative expansion in powers of momenta is done,
leading to chiral perturbation theory ($\chi PT$) which has allowed great progress in
hadron physics. A very important development in this direction is done with
the introduction of unitarity in coupled channels which has allowed to extend
the predictions of chiral theory at higher energies than is possible with
chiral perturbation theory. A review of the basic ideas can be found in 
\cite{review}. One of the important issues of this unitarized theory is that it
allows to deal with resonances, some of which appear necessarily as a
consequence of the dynamics of the chiral Lagrangians, in a similar way as
bound states or resonances appear in potential theory in Quantum Mechanics.
 We call these resonances
dynamically generated  and by now about 10-15 percent of the
particles in the PDG can qualify for this nature, meaning that the meson- meson
cloud, or meson-baryon cloud, becomes the basic part of the wave function
overriding the relevance of the primary building blocks: constituent 
$q \bar{q}$ or three $q$ for the case of
 mesons or baryons, respectively.  I shall mention here some of these
cases.  

  The findings about the nature of some resonances have also consequences in
  nuclear physics, since basic features associated to the chiral dynamics of
  the resonances show up in interesting ways in nuclear processes which can
  then be used to stress the nature of these resonances.  I shall also give
  some example of that in this talk.  A more extensive description of the ideas
  exposed here can be found in \cite{puri}, but we concentrate here on the
  recent developments not discussed in \cite{puri}.
  
\section{The meson baryon interaction} 

  We skip here details on the chiral lagrangians which can be found in
  \cite{review,puri,ulf,ecker} and sketchhow unitarity enters the
  framework. 
One can find a systematic and easily comprehensible derivation 
 of the  ideas of the N/D method applied for the first time to the meson baryon system in
 \cite{Oller:2000fj}, which we reproduce here below and which follows closely
 the similar developments used before in the meson meson interaction \cite{nsd}.
 One defines the transition $T-$matrix as $T_{i,j}$ between the coupled channels which couple to
 certain quantum numbers. For instance in the case of  $\bar{K} N$ scattering studied in
 \cite{Oller:2000fj} the channels with zero charge are $K^- p$, $\bar{K^0} n$, $\pi^0 \Sigma^0$,$\pi^+
 \Sigma^-$, $\pi^- \Sigma^+$, $\pi^0 \Lambda$, $\eta \Lambda$, $\eta \Sigma^0$, 
 $K^+ \Xi^-$, $K^0 \Xi^0$.
 Unitarity in coupled channels is written as
 
\begin{equation} 
Im T_{i,j} = T_{i,l} \rho_l T^*_{l,j}
\end{equation}
where $\rho_i \equiv 2M_l q_i/(8\pi W)$, with $q_i$  the modulus of the c.m. 
three--momentum, and the subscripts $i$ and $j$ refer to the physical channels. 
 This equation is most efficiently written in terms of the inverse amplitude as
\begin{equation}
\label{uni}
\hbox{Im}~T^{-1}(W)_{ij}=-\rho(W)_i \delta_{ij}~,
\end{equation}
The unitarity relation in Eq. (\ref{uni}) gives rise to a cut in the
$T$--matrix of partial wave amplitudes, which is usually called the unitarity or right--hand 
cut. Hence one can write down a dispersion relation for $T^{-1}(W)$ 
\begin{equation}
\label{dis}
T^{-1}(W)_{ij}=-\delta_{ij}\left\{\widetilde{a}_i(s_0)+ 
\frac{s-s_0}{\pi}\int_{s_{i}}^\infty ds' 
\frac{\rho(s')_i}{(s'-s)(s'-s_0)}\right\}+{\mathcal{T}}^{-1}(W)_{ij} ~,
\end{equation}
where $s_i$ is the value of the $s$ variable at the threshold of channel $i$ and 
${\mathcal{T}}^{-1}(W)_{ij}$ indicates other contributions coming from local and 
pole terms, as well as crossed channel dynamics but {\it without} 
right--hand cut. These extra terms
are taken directly from $\chi PT$ 
after requiring the {\em matching} of the general result to the $\chi PT$ expressions. 
Notice also that the curled bracket, that we call the $g(s)$ function, 
is the familiar scalar loop integral.

One can further simplify the notation by employing a matrix formalism. 
Introducing the 
matrices $g(s)={\rm diag}~(g(s)_i)$, $T$ and ${\mathcal{T}}$, the latter defined in 
terms 
of the matrix elements $T_{ij}$ and ${\mathcal{T}}_{ij}$, the $T$-matrix can be written as:
\begin{equation}
\label{t}
T(W)=\left[I-{\mathcal{T}}(W)\cdot g(s) \right]^{-1}\cdot {\mathcal{T}}(W)~.
\end{equation}
which can be recast in a more familiar form as 
 \begin{equation}
\label{ta}
T(W)={\mathcal{T}}(W)+{\mathcal{T}}(W) g(s) T(W)
\end{equation}
Now imagine one is taking the lowest order chiral amplitude for the kernel
${\mathcal{T}}$ as done in
\cite{Oller:2000fj}. Then the former equation is nothing but the Bethe Salpeter equation with the
kernel taken from the lowest order Lagrangian and  factorized  on  shell, the same
approach followed in \cite{kaon}, where different arguments were used to justify the on shell
factorization of the kernel.

The on shell factorization of the kernel, justified here with the N/D method,
renders the set of coupled Bethe Salpeter integral equations a simple set of
algebraic equations.

\section{Poles of the T-matrix}

\begin{figure}
\begin{center}
\includegraphics[scale=0.4]{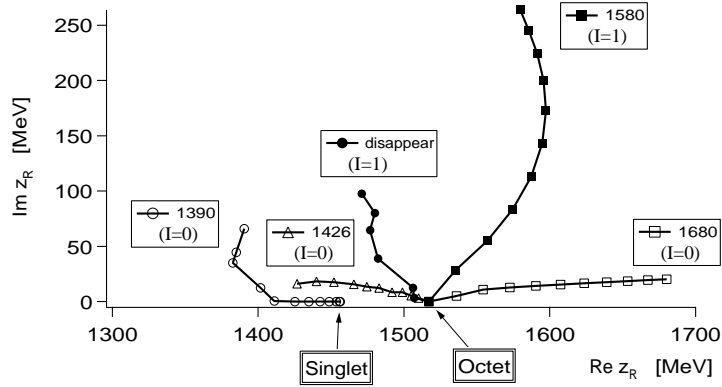}
 \caption{Trajectories of the poles in the scattering amplitudes obtained by
  changing the SU(3) breaking gradually. 
  The symbols correspond to the step size $\delta x =0.1$ with $x$ running from 0 to 1 
  . The results are from
  \protect\cite{Jido:2003cb}.}
  \label{fig:tracepole}
\end{center}  
\end{figure}

  The amplitudes of the channels discussed above are extrapolated to the
  complex
   second Riemann sheet and we search for poles  there.  Here I
   discuss  recent finding concerning the two poles corresponding to the 
 $\Lambda(1405)$ resonance.  In fig.~\ref{fig:tracepole} we can see, 
 from the work of \cite{Jido:2003cb}, the
 trajectories of the poles when we start from an SU(3) symmetric situation and
 go to
 the real world where this symmetry is broken.  This appears because of the 
 different masses
 of the mesons and the baryon in the study of the interaction of the octet of the pion 
 and the octet of the
 nucleon respectively. The SU(3) symmetric case corresponds to the poles in the
 real axis where we have two degenerate octets and a singlet. As soon as SU(3)
 symmetry starts being broken the degeneracy of the two octets disappears and
 two branches for I=0 and for I=1 develop which, at the end of the trajectories that
 represent the real world, correspond to well know resonances like the
 $\Lambda(1670)$ etc.

\begin{figure}[htb]
 \begin{center} 
\includegraphics[scale=0.4]{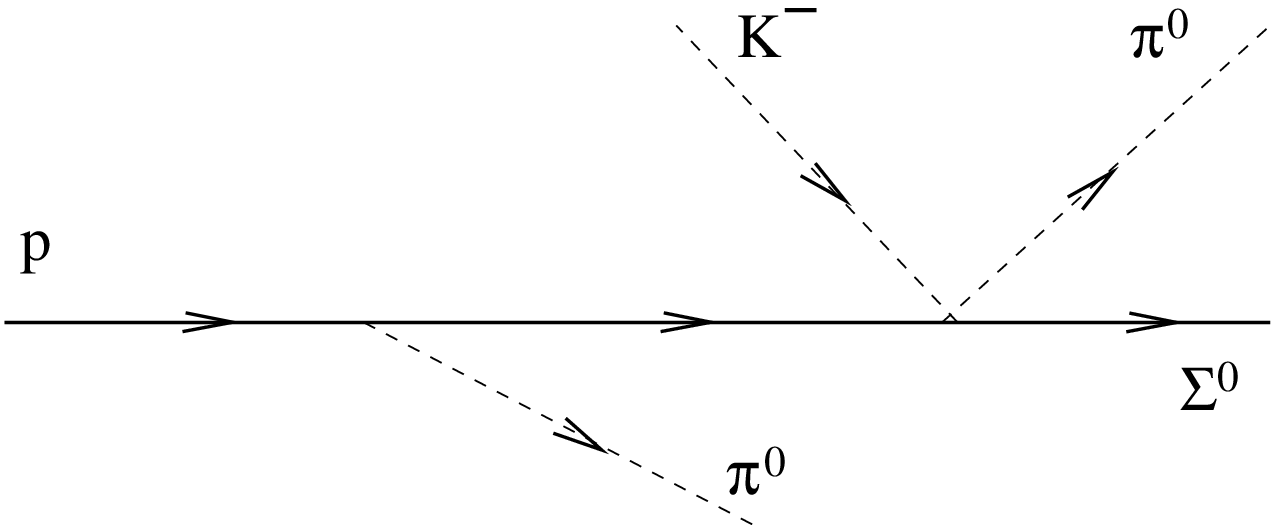}  
\includegraphics[scale=0.4]{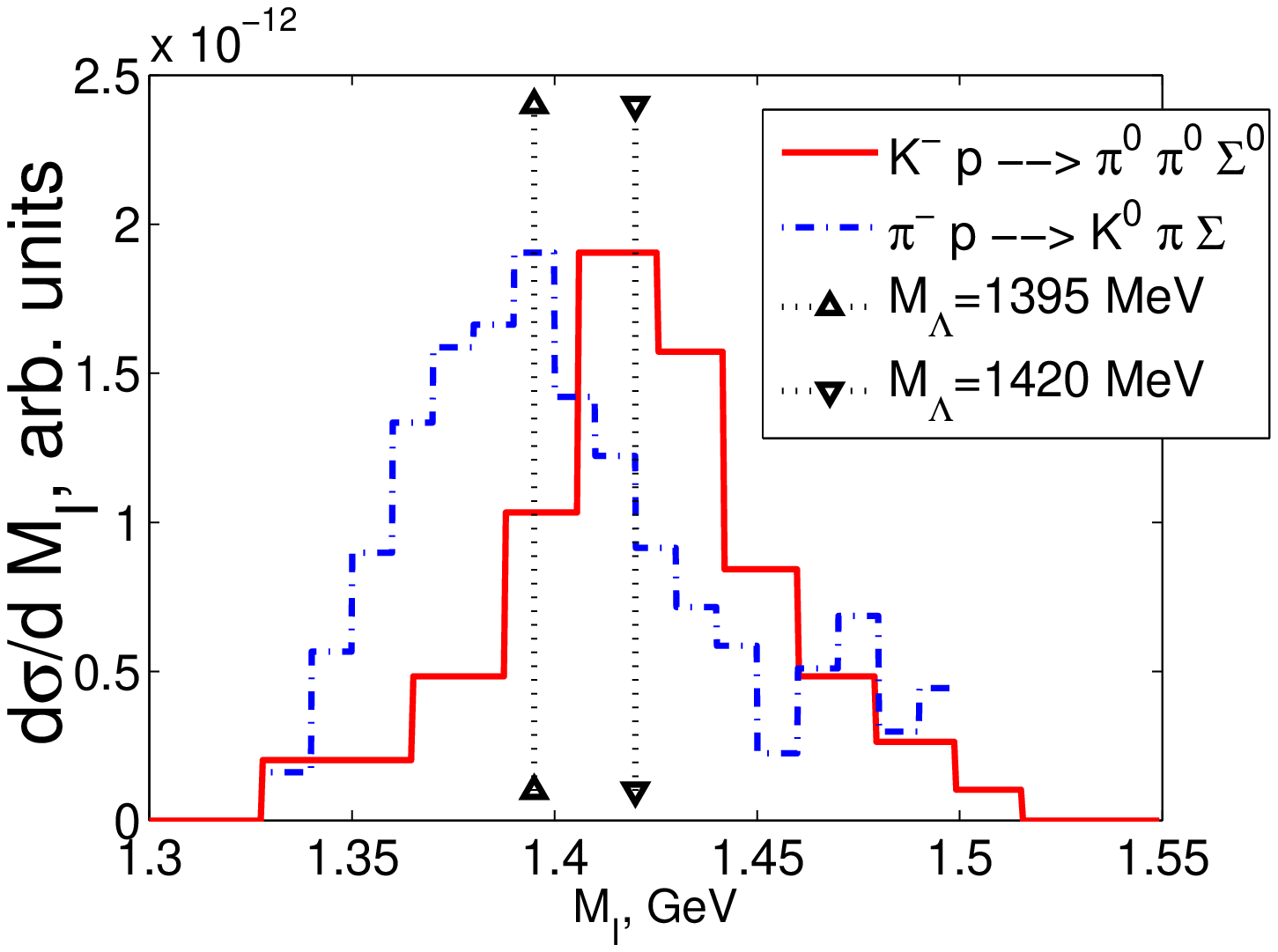}
 \caption{Nucleon pole term for the $K^- p \to \pi^0 \Sigma^0$ (left) and 
  two experimental shapes of  $\Lambda(1405)$ resonance (right). 
 See text for more details.} 
 \label{two_exp}
 \end{center}
\end{figure}

The interesting thing to note is that in the region of the $\Lambda(1405)$
   there are two poles, one from the original singlet and another one from 
   one of the branches of the original octet. In practice it is impossible to
   see two peaks in any reaction because the two resonances overlap,
    but the fact that the two poles couple differently to $\bar{K}N$ and $\pi
    \Sigma$ (the narrower pole couples strongly to $\bar{K} N$ and the wider
    one to $\pi \Sigma$)  has as a consequence that the $\Lambda(1405)$ should show up
    with quite different shapes in different reactions depending on their
    particular dynamics.  This is indeed the case as
    could be demonstrated recently with the performance of the 
    $K^- p \to \pi^0 \pi^0 \Sigma^0$  \cite{Prakhov} experiment. This reaction has been
    studied in \cite{Magas:2005vu} and the basic mechanism for the reaction is
    given by fig.~\ref{two_exp} left,
    where we see that it is dominated by the $K^- p \to \pi^0 \Sigma^0 $ amplitude
which has a strong contribution from the narrow pole at higher energies. We
should then expect that the peak of the $\Lambda(1405)$ appears at higher
energies in this reaction and with a smaller width, and this is the case as one
can see in fig. ~\ref{two_exp} right  by comparing with the standard 
$\Lambda(1405)$ extracted from
the  $\pi^- p \to K^0 \pi \Sigma$ reaction which was studied in \cite{hyodo}.

  The two experimental figures for the shapes of the $\Lambda(1405)$ in both
  reactions can be seen in Fig.~\ref{two_exp} right, giving a strong support to the
  existence of the two $\Lambda(1405)$ states.

\vspace{-0.7cm}
  
\section{The interaction of the decuplet of baryons with the octet of mesons}
Given the success of the chiral unitary approach in generating dynamically low
energy resonances from the interaction of the octets of stable baryons
and the pseudoscalar mesons,   in 
\cite{Kolomeitsev:2003kt} the
interaction of the decuplet of $3/2^+$ with the octet of pseudoscalar mesons 
was studied and shown to
lead to many states which were associated to experimentally well 
established $3/2^-$ resonances. One of these would correspond to the
$\Lambda(1520)$.

\begin{figure}[h]
\begin{center}
\includegraphics[scale=0.3]{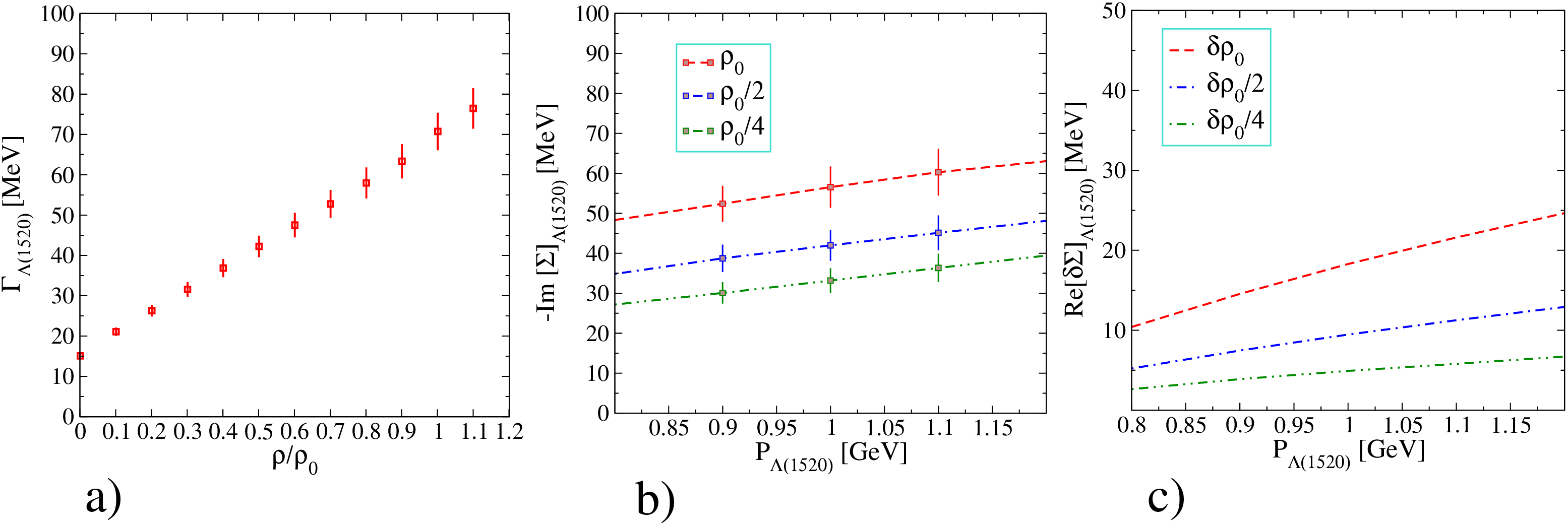}
\includegraphics[scale=0.3]{fig7.eps}
 \caption{Cross section for the $K^- p \to \pi^0 \pi^0 \Lambda$ (left) and 
$K^-p \to \pi^+ \pi^- \Lambda$ \cite{Mast:1973gb} right.}
\label{sigfig2}
\end{center}
\end{figure}

In \cite{sarkar,Roca:2006sz} a refinement of the approach discussed above has been done
including the $\bar{K} N$ and $\pi \Sigma$ decay channels of the $\Lambda(1520)$ and
finetuning the subtraction constants in the $g$ function, such that a good
agreement with the position and width of the $\Lambda(1520)$ is attained.  With this
new information one can face the study of the reaction $K^-p \to
\pi^0\pi^0\Lambda$ by
using the mechanisms of $K^- p \to \pi^0 \Sigma^{*0} \to \pi^0 \pi^0 \Lambda$ 
which provides the dominant contribution to
the reaction at energies close to the $\Lambda(1520)$.  At higher energies of the
experiment of \cite{Prakhov:2004ri} one finds other background mechanisms 
(dashed line in fig. ~\ref{sigfig2}) providing
a contribution that helps bring the theory and experiment in good agreement, as
we can see in fig. ~\ref{sigfig2}.  One can see in the figure that up to 575 MeV/c of
momentum of the $K^-$ the mechanism based on the strong coupling of the  
$\Lambda(1520)$ to the $\pi \Sigma(1385)$ channel is largely dominant and provides
the right strength of the cross section.  Obviously it would be most interesting
to investigate the region of lower energies of the $K^-$ in order to see if
the predictions done by the theory are accurate. This could be done for a
different reaction, the $K^-p \to \pi^+ \pi^- \Lambda$ \cite{Mast:1973gb}, where
the cross section is double than for $K^- p \to \pi^0 \pi^0 \Lambda$. As we can
see in fig.~ \ref{sigfig2}, the predictions of the theory are in good agreement
with the data.

\vspace{-0.7cm}

\section{$\bar{K}$ in a nuclear medium}
The $\bar{K}$ interaction in a nuclear medium has been the subject of many
studies. In \cite{Ramos:1999ku} a selfconsistent calculation of the $\bar{K}$
nucleus optical potential is done and the results are  consistent
with data of kaonic atoms as shown in \cite{zaki}.  This potential has a strength of
about 50 MeV attraction at normal nuclear matter and an imaginary part of
about 50 MeV.  This leads to deeply bound states in nuclei of about 30 MeV
binding and widths of around 100 MeV.  It was hence surprising to find other
potentials where, with only three nucleons, the $\bar{K}$ potential reaches a
strength of about 650 MeV at the center for the nucleus \cite{Akaishi:2005sn}. 
A closer inspection shows that the huge results come from several
approximations among them the lack of selfconsistency in the calculations that,
according to \cite{lutz}, leads to meaningless results, plus the shrinking of the
nucleus to attain ten times nuclear matter density at the center of the nucleus. 
  Based on this potential, some peaks seen at KEK in the proton spectra
  following the absorption of $K^-$ on $^4 He$ were interpreted as a signal of
  deeply bound kaonic states.  Recently, it was shown in  \cite{osettoki} that the
  peaks of the experiments could be easily interpreted in terms of $K^-$
  absorption by pairs of nucleons leading to $\Sigma p$ or $\Lambda p$ without
  further interaction of the baryons with the nucleus.  This claim has been
  recently reinforced by an experiment at FINUDA  \cite{Agnello:2006gx} 
  where the proton spectra is
  also studied following $K^-$ absorption in different nuclei and the same
  peaks are seen. An additional measurement of pions in coincidence shows that
  they come from the decay of the $\Sigma$ for the peak at lower proton 
  momentum, hence, confirming the claims of \cite{osettoki}.
   
    The FINUDA collaboration has another paper \cite{Agnello:2005qj} in which a
    wider peak has been seen in the invariant mass of $p \Lambda$ measured 
    back to back
    following $K^-$ absorption in different nuclei and which was 
    claimed to correspond the a $K^- pp$ bound
    state.  In another recent paper \cite{Magas:2006fn} 
    it has been shown that such a peak appears
 unavoidably as a consequence of the mechanism of  $K^-$ absorption by a pair of
 nucleons going to $p \Lambda$, followed by rescattering of the $p$ or the
 $\Lambda$ in the nucleus.  Thus, the claims made for deeply bound kaons in
 nuclei with binding energies of about 200 MeV and narrow widths are unfounded
 and this was the conclusion of a recent meeting explicitly devoted to this
 issue in $ECT^*$ Trento \cite{trento}.
 
 \section{$\Lambda(1520)$ in a nuclear medium}
 Within these techniques there is work done in $\eta$ selfenergy in the
 medium, $\rho$ and $\phi$ renormalization in nuclei and others, but I shall
 concentrate here on the latest issue which has connections with section 4,
 where we studied the interaction of the decupled of baryons with the octet
 of nucleons.  The $\Lambda(1520)$ is one of the
 resonances dynamically generated with the channels $\pi \Sigma(1385)$ and $K
 \Xi$. Actually, this picture is only qualitative since the $\Lambda(1520)$
 couples also to the $\pi \Sigma$ and $\bar{K} N$ channels to which it decays. 
 The study done in \cite{Roca:2006sz} including these channels 
 finds that the coupling of the resonance to the 
 $\pi \Sigma(1385)$ is the strongest of all, as if it kept memory of what would
 happen in a simplified ideal world.  
 
 \begin{figure}[h]
 \begin{center}
\includegraphics[scale=0.2]{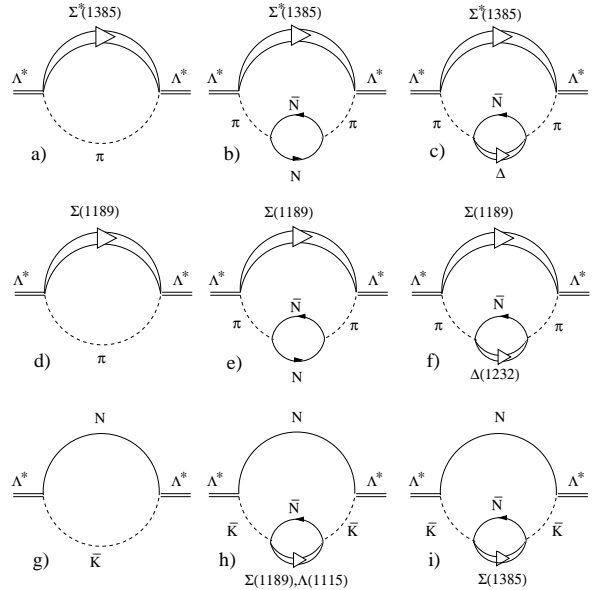}
\caption{Diagrams for the $\Lambda(1520)$ selfnergy in the nuclear medium}
\label{diagrams}
\end{center}
\end{figure}
 
\begin{figure}[h]
\begin{center}
\includegraphics[scale=0.3]{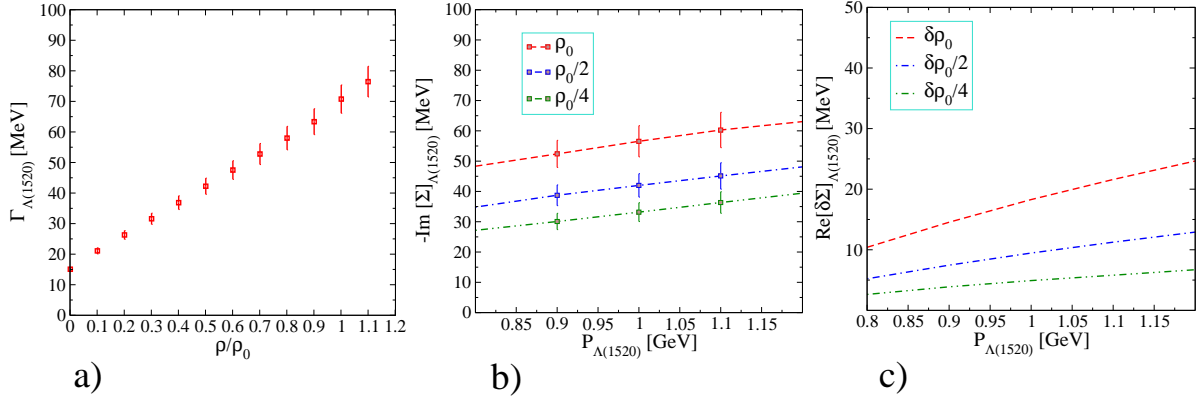}
\caption{Results for the $\Lambda(1520)$ selfenergy as a function of the nuclear
density and $\Lambda(1520)$ momentum.}
\label{resultslam}
\end{center}
\end{figure}

 Even if the $\Lambda(1520)$ couples strongly to the $\pi \Sigma(1385)$ 
 channel, it does not decay into this state because it is below the threshold
 of $\pi \Sigma(1385)$.  However, in a nucleus the pion can easily excite 
 a $ph$ which has energy starting from zero.
 Thus we gain 140 MeV phase space for the decay and this leads to a width in
 the medium from only this channel which is already bigger than the 14 MeV free
 width of the $\Lambda(1520)$. This is not the only source of medium modification for
 the resonance, since the other decay channels are also modified in the medium.
 A detailed study has been done \cite{Kaskulov:2005uw} (see diagrams in 
 fig.~\ref{diagrams} ) and we present the results in fig. ~
 \ref{resultslam}.

\vspace{-0.5cm}

\section{Conclusions}
The use of chiral Lagrangians for the meson baryon interaction and the unitary
extensions of chiral perturbation theory have allowed to face a large amount of
problems which were barred to standard perturbation techniques. It has
opened the door to the study of many baryonic resonances which qualify neatly as
dynamically generated resonances, or quasibound states of meson baryon. Thanks
to this, a quantitative description of the meson baryon interaction at 
intermediate energies is now possible and with this, an important systematics has
been introduced in the many body problem to face issues on the renormalization of
hadron properties in the nuclear medium. The constructed scheme is rather powerful
and allows to make predictions which consecutive experiments are proving right.
Two of these predictions, the two states for the $\Lambda(1405)$, and the nature
of the $\Lambda(1520)$ roughly as a quasibound state of $\pi \Sigma(1385)$, have found
strong support from two very recent experiments. The spectacular change of the
width of the $\Lambda(1520)$ in the medium is partly due to the strong coupling
of the resonance to the $\pi \Sigma(1385)$ as predicted by chiral unitary
dynamics. Finally, we also discussed that chiral theory, implementing
selfconsistency in the many body calculation, leads to a moderate attraction
for $\bar{K}$ in  nuclei, and that recents claims for deeply bound kaon
states, which would require a huge $\bar{K}$ optical potential, were a
consequence of an incorrect identification of some peaks for which a natural
conventional explanation has been just found. 

 Altogether, one is seeing through all
this work that chiral dynamics is a key ingredient that allows a unified
description of much of the hadronic world at low and intermediate energies.

\section{Acknowledgments}
This work is partly supported by the Spanish CSIC and JSPS collaboration, the
 DGICYT contract number BFM2003-00856,
and the E.U. EURIDICE network contract no. HPRN-CT-2002-00311. 
This research is part of the EU
      Integrated Infrastructure Initiative
      Hadron Physics Project under contract number
      RII3-CT-2004-506078.


\begin{thebibliography}{99}
\bibitem{review}
  J.~A.~Oller, E.~Oset and A.~Ramos,
   Prog.\ Part.\ Nucl.\ Phys.\  {\bf 45} (2000) 157.
  
\bibitem{puri}
  E.~Oset, D.~Cabrera, V.~K.~Magas, L.~Roca, S.~Sarkar, M.~J.~Vicente Vacas and A.~Ramos,
  Lectures at Puri Hadron Workshop, Pramana {\bf 66} (2006) 731
  [arXiv:nucl-th/0504033].
  
  \bibitem{ulf}U. G. Meissner, Rep. Prog. Phys. {{56}} (1993)
903; V. Bernard, N. Kaiser and U. G. Meissner, Int. J. Mod. Phys. {{E4}} (1995)
193.

\bibitem{ecker}G. Ecker, Prog. Part. Nucl. Phys. {{35}} (1995) 1.

\bibitem{Oller:2000fj}
J.~A.~Oller and U.~G.~Meissner,
Phys.\ Lett.\ B {\bf 500} (2001) 263


\bibitem{nsd} J. A. Oller and E. Oset, Phys. Rev. D {\bf 60} (1999) 074023.

\bibitem{kaon}
  E.~Oset and A.~Ramos,
  Nucl.\ Phys.\ A {\bf 635} (1998) 99.
 

\bibitem{Jido:2003cb}
D. Jido, J.A. Oller, E. Oset, A. Ramos, U.G. Meissner,
Nucl.\ Phys.\ A {\bf 725} (2003) 181.

\bibitem{Prakhov}
S.~Prakhov {\it et al.}  [Crystall Ball Collaboration],
Phys.\ Rev.\ C {\bf 70} (2004) 034605.

\bibitem{Magas:2005vu}
  V.~K.~Magas, E.~Oset and A.~Ramos,
  Phys.\ Rev.\ Lett.\  {\bf 95} (2005) 052301.

  
\bibitem{hyodo}
  T.~Hyodo, A.~Hosaka, E.~Oset, A.~Ramos and M.~J.~Vicente Vacas,
  Phys.\ Rev.\ C {\bf 68} (2003) 065203.
  
  
\bibitem{Kolomeitsev:2003kt}
E.~E.~Kolomeitsev and M.~F.~M.~Lutz,
Phys.\ Lett.\ B {\bf 585} (2004) 243.

\bibitem{sarkar}
  S.~Sarkar, E.~Oset and M.~J.~Vicente Vacas,
  Nucl.\ Phys.\ A {\bf 750} (2005) 294.
  
  
\bibitem{Roca:2006sz}
  L.~Roca, S.~Sarkar, V.~K.~Magas and E.~Oset,
  Phys.\ Rev.\ C {\bf 73} (2006) 045208.
  
\bibitem{Prakhov:2004ri}
  S.~Prakhov {\it et al.},
  Phys.\ Rev.\ C {\bf 69} (2004) 042202.
  
  
\bibitem{Mast:1973gb}
  T.~S.~Mast, M.~Alston-Garnjost, R.~O.~Bangerter, A.~Barbaro-Galtieri, F.~T.~Solmitz and R.~D.~Tripp,
   Phys.\ Rev.\ D {\bf 7} (1973) 5.
  
  
\bibitem{Ramos:1999ku}
  A.~Ramos and E.~Oset,
  Nucl.\ Phys.\ A {\bf 671} (2000) 481.
  
\bibitem{zaki} S. Hirenzaki, Y.Okumura, H. Toki, E. Oset and A. Ramos, Phys.
Rev. C61 (2000) 055205.  

\bibitem{Akaishi:2005sn}
  Y.~Akaishi, A.~Dote and T.~Yamazaki,
  Phys.\ Lett.\ B {\bf 613} (2005) 140.
  
 \bibitem{lutz} M. Lutz, Phys. Lett. B426 (1998) 12. 
 
\bibitem{osettoki}
  E.~Oset and H.~Toki, Phys. Rev. C, in print.
  arXiv:nucl-th/0509048.

\bibitem{Agnello:2006gx}
  M.~Agnello {\it et al.},
  arXiv:nucl-ex/0606021.
  
\bibitem{Agnello:2005qj}
  M.~Agnello {\it et al.}  [FINUDA Collaboration],
    Phys.\ Rev.\ Lett.\  {\bf 94} (2005) 212303.
  
\bibitem{Magas:2006fn}
  V.~K.~Magas, E.~Oset, A.~Ramos and H.~Toki,
  arXiv:nucl-th/0601013.
  
 \bibitem{trento} http://www.itkp.uni-bonn.de/percent7Erusetsky/TRENTO06/trento06.html
 

  
\bibitem{Kaskulov:2005uw}
  M.~Kaskulov and E.~Oset,
  Phys.\ Rev.\ C {\bf 73} (2006) 045213.
  
  

\end{thebibliography}
\end{document}